\documentclass[showpacs,preprintnumbers,amsmath,amssymb,twocolumn,superscriptaddress,prb]{revtex4}
\usepackage{graphicx}
\usepackage{amsfonts}
\usepackage{amsmath, amsthm, amssymb}
\usepackage{dsfont}
\usepackage{times}

\newcommand{\ve}[1]{{\boldsymbol{#1}}}
\newcommand{\vk}{\ve{k}} % Vector k
\newcommand{\e}[1]{\mathrm{e}^{#1}}
\newcommand{\eg}{\textit{e.g. }}%[syn: f.eks., for example, for instance]
\newcommand{\etal}{\emph{et al.}}
\def\i{\mathrm{i}}

\begin{document}
\title[Anomalous Finite Size Effects on Surface States in the Topological Insulator Bi$_2$Se$_3$ ]{Anomalous Finite Size Effects 
on Surface States in the Topological Insulator Bi$_2$Se$_3$}
\author{Jacob Linder}
\affiliation{Department of Physics, Norwegian University of
Science and Technology, N-7491 Trondheim, Norway}
\author{Takehito Yokoyama}
\affiliation{Department of Applied Physics, University of Tokyo, Tokyo 113-8656, Japan}
\author{Asle Sudb{\o}}
\affiliation{Department of Physics, Norwegian University of
Science and Technology, N-7491 Trondheim, Norway}

\date{Received \today}
\begin{abstract}
\noindent We study how the surface states in the strong topological insulator Bi$_2$Se$_3$ are influenced by finite size effects, and 
compare our results with those recently obtained for 2D topological insulator HgTe. We demonstrate two important distinctions: 
\textit{(i)} contrary to HgTe, the surface-states in Bi$_2$Se$_3$ display a remarkable robustness towards decreasing the width 
$L$ down to a few nm, thus ensuring that the topological surface states remain intact, and \textit{(ii)} the gapping due to the 
hybridization of the surface 
states features  an oscillating exponential decay as a function of $L$ in Bi$_2$Se$_3$ in sharp contrast to HgTe. Our findings 
suggest that Bi$_2$Se$_3$ is suitable for nanoscale applications in quantum computing or spintronics. Also, we propose a way to 
experimentally detect both of the predicted effects.
\end{abstract}

\maketitle

\section{Introduction}

A new state of matter, known as the topologically insulating state, has recently been experimentally observed \cite{konig_jpsj_08, konig_science_07,hsieh} after the successful prediction of its existence in HgTe \cite{bernevig_science_06}. This state is characterized by the topological protection of the conducting states that form at the edges (in 2D) \cite{kane_prl_05, murukami_prl_06, bernevig_prl_06} or surfaces 
(in 3D) \cite{Fu3D,Qi} of such materials, whereas the bulk states remain insulating due to a charge excitation gap. This 
distinguishes topological insulators from conventional insulators that do not feature such edge/surface states, and introduces 
several interesting effects\cite{Qi2,Yokoyama,Tanaka,mondal_arxiv_09}. 
In particular, topological insulators could find use in quantum computation since the topologically 
protected edge/surface states remain insensitive to disorder \cite{Fu1,Fu2,Beenakker,Beenakker2}.

It has recently been realized that Bi$_2$Se$_3$ is a three-dimensional topological insulator with a large charge excitation gap in the 
bulk \cite{zhang_nphys_09}. The surface states have an energy dispersion that is linear in momentum and thus form a Dirac cone at low 
energy, similarly to graphene. In stark contrast, however, Kramer's theorem does not guarantee the survival of the edge states in 
graphene in the presence of perturbations since it holds an even number of Dirac cones inside the Fermi contour.\cite{Fu3D} In Bi$_2$Se$_3$, the 
number of Dirac points inside the Fermi arc is odd which activates the protection from Kramer's degeneracy theorem. This has very recently been experimentally observed \cite{xia_nphys_09}.

The prospect of utilizing the protected surface states in topological insulators such as Bi$_2$Se$_3$ in actual devices related to 
quantum computing or spintronics demands that finite size effects are taken seriously. This fact is underlined by the finding of 
Ref. \cite{zhou_prl_08} which showed that the edge states of HgTe quantum wells become gapped due to a finite-size effect as the 
width $L$ is decreased. The gapping was shown to become experimentally measurable around $L\simeq200$ nm, suggesting that the 
material loses its exotic edge state properties in this region. Clearly, this places severe restrictions on potential 
use of HgTe quantum wells in applications on the nm-scale.

\begin{figure}[t!]
\centering
\resizebox{0.40\textwidth}{!}{
\includegraphics{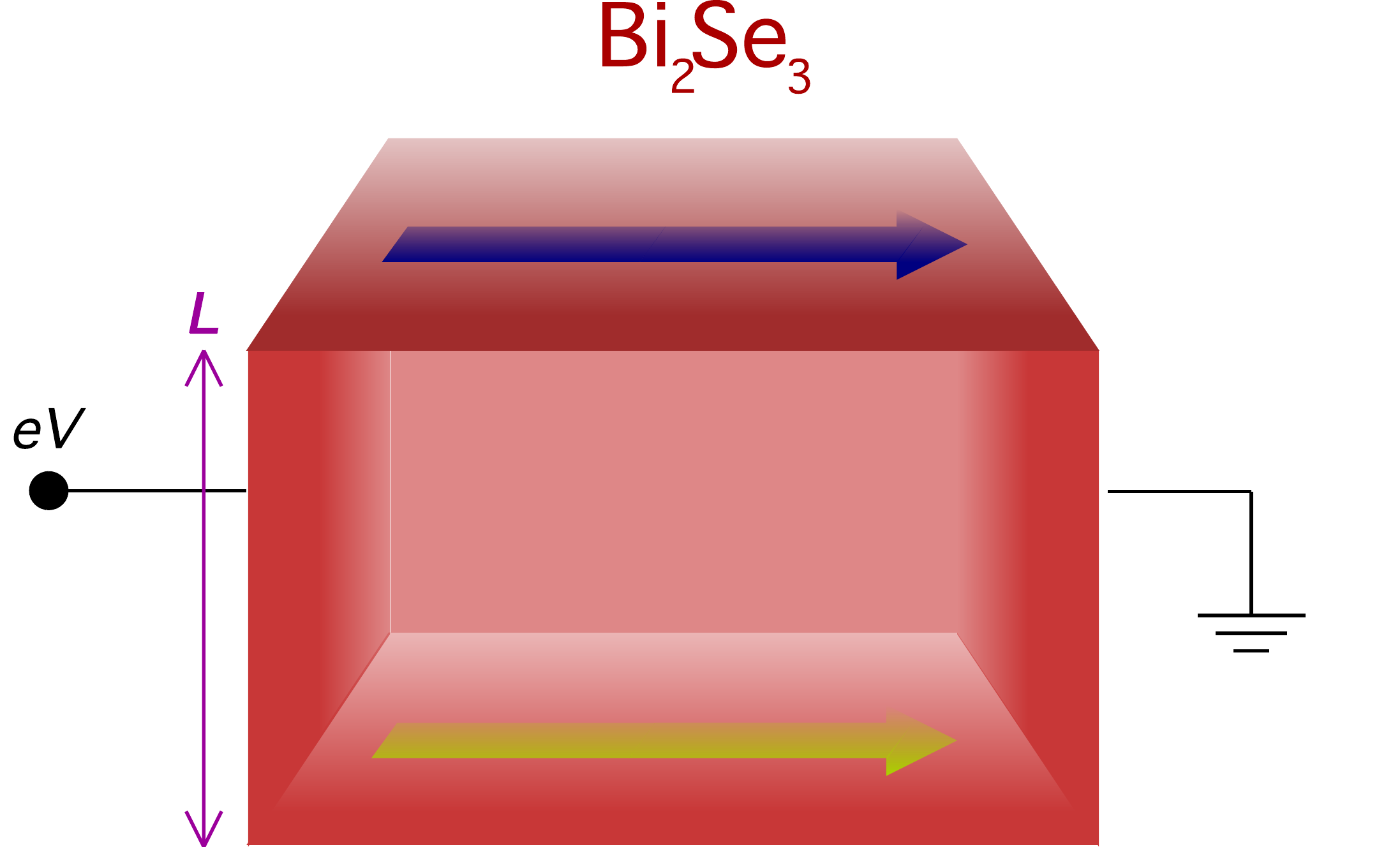}}
\caption{(Color online) Suggested experimental setup: two-terminal geometry with the topological insulator Bi$_2$Se$_3$. The direction 
of the current carried by the surface states is shown by the arrows, and the width of the sample is $L$. For $L \sim 2 \hbar v_F/|M|$, 
finite-size effects become important. Here, $|M|$ is the bulk charge excitation gap. }
\label{fig:model} 
\end{figure}

In this work, we will demonstrate how the situation changes dramatically when considering Bi$_2$Se$_3$. Within a combined 
analytical 
and numerical approach, we show how the surface states in Bi$_2$Se$_3$ display a remarkable robustness towards finite-size effects, 
becoming measurably gapped only when the width drops to a few nm. This means that samples of Bi$_2$Se$_3$ can be made several tens 
of times smaller than HgTe while still retaining their characteristic surface states giving rise to the quantum spin Hall 
effect. We explain this observation in terms of the large charge excitation gap $|M|$ in Bi$_2$Se$_3$ which gives rise to 
a very short localization length of the surface states. Moreover, we show how the finite-size induced gapping in the surface 
states displays a qualitatively different dependence on $L$ in Bi$_2$Se$_3$ compared to HgTe, namely an oscillating exponential 
decay. Both of these effects can be measured in a two-terminal geometry as sketched in Fig. \ref{fig:model}.

\begin{figure}
\centering
\resizebox{0.5\textwidth}{!}{
\includegraphics{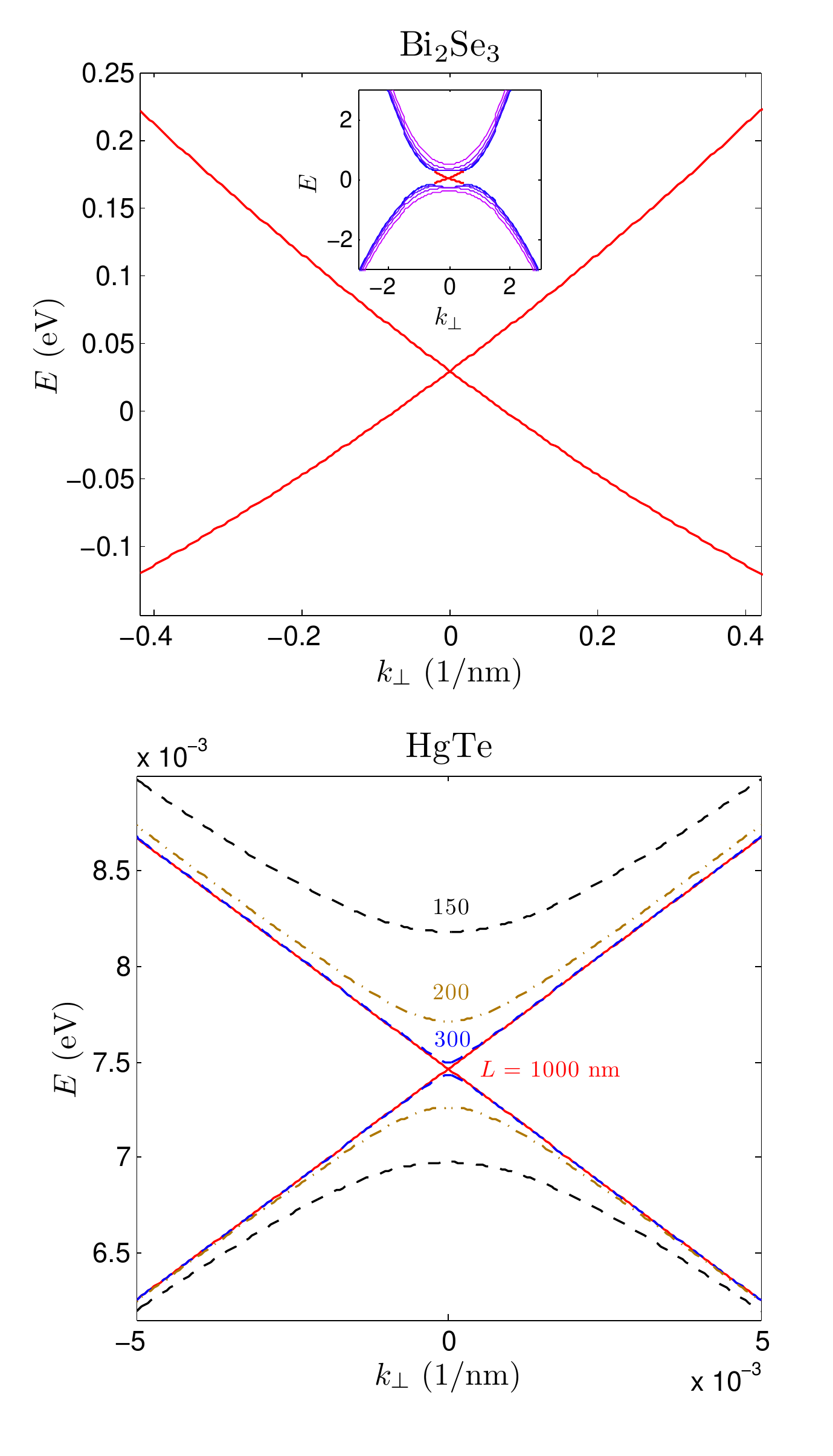}}
\caption{(Color online) Plot of the surface/edge state energy dispersions versus the transverse momentum $k_\perp$ in the case of 
Bi$_2$Se$_3$ and HgTe. For Bi$_2$Se$_3$, we give only the result for $L=1000$ nm since the dispersion remains completely unchanged 
down to $L=150$ nm, in contrast to HgTe where a gap opens up. The inset of Bi$_2$Se$_3$ shows the bulk bands and the surface states 
with a gap of $2|M|=0.56$ eV opening at the $\Gamma$ point. }
\label{fig:energyvsk} 
\end{figure}

\section{Theory}

The effective low-energy Hamiltonian for Bi$_2$Se$_3$ centered around the $\Gamma$ point in the Brillouin Zone may 
be written as \cite{zhang_nphys_09}:
\begin{align}\label{eq:H}
\hat{\mathcal{H}} &= \begin{pmatrix}
\varepsilon_\vk \underline{1} + \mathcal{M}_\vk\underline{\tau_z} + A_1k_z\underline{\tau_x} & A_2k_-\underline{\tau_x}\\
A_2k_+\underline{\tau_x} & \varepsilon_\vk\underline{1} + \mathcal{M}_\vk \underline{\tau_z} - A_1k_z \underline{\tau_x} \\
\end{pmatrix},
\end{align}
where we have defined the following quantities: 
\begin{align}
\varepsilon_\vk = C + D_1k_z^2 + D_2k_\perp^2,,\notag\\ \mathcal{M}_\vk = M - B_1k_z^2 - B_2k_\perp^2,
\end{align} 
and $k_\pm = k_x\pm \i k_y$.
Here, $\tau_z$ and $\tau_x$ are the Pauli matrices in standard notation,
while the parameters $\{A_j,B_j,C,D_j,M\}$ describe the band-structure in Bi$_2$Se$_3$, obtainable by first-principles calculations. Such a fitting procedure was undertaken in Ref. \cite{zhang_nphys_09}, with the result $C = -6.8\times10^{-3} \text{ eV},$ $M = 0.28 \text{ eV}$, and
\begin{align}\label{eq:parameters}
A_1 &= 2.2 \text{ eV$\cdot$\AA},\; A_2 = 4.1 \text{ eV$\cdot$\AA},\; B_1 = 10 \text{ eV$\cdot$\AA$^2$},\notag\\
 B_2 &= 56.6 \text{ eV$\cdot$\AA$^2$},\; D_1 = 1.3 \text{ eV$\cdot$\AA$^2$},\; D_2 = 19.6 \text{ eV$\cdot$\AA$^2$}.
\end{align}
It should be noted that there is an anisotropy along the $\hat{\boldsymbol{z}}$-axis 
and that the full 3D structure of Bi$_2$Se$_3$ has been taken into account. The basis $\Psi$ we have used for the $4\times4$ Hamiltonian is 
\begin{align}
\Psi = (|P1_z^+,\uparrow\rangle, |P2_z^-,\uparrow\rangle, |P1_z^+,\downarrow\rangle, |P2_z^-,\downarrow\rangle)^\text{tr},
\end{align}
where tr denotes the transpose operation. The above $p$-orbital states are the relevant ones near the Fermi level of Bi$_2$Se$_3$ and 
may be classified according to their parity $\pm$ since inversion symmetry is preserved on the lattice. The states $|P1\rangle$ stem 
from the Bi atoms, whereas $|P2\rangle$ stems from the Se atoms. The $|P1_z^+,\sigma\rangle$ and $|P2_z^-,\sigma\rangle$ states ($\sigma=\uparrow, \downarrow$) have opposite parity and the order of them near the Fermi level is interchanged when spin-orbit coupling is taken into account \cite{zhang_nphys_09}. In this way, the spin-orbit coupling effect is responsible for driving the system into a topologically insulating phase.

It is interesting to observe that the Hamiltonian Eq. (\ref{eq:H}) 
contains spin-mixing terms due to the off-diagonal entries $A_2k_\pm\underline{\tau_x}$. This in contrast to the 2D topological 
insulator HgTe, which is diagonal in spin-space. As a result, one might expect new features in the spin-current of Bi$_2$Se$_3$ 
carried by the topological surface states. 

\section{Results and discussion}

 We will consider a finite width $L$ in the $z$-direction with open boundary conditions at the 
edges, i.e. 
\begin{align}
\Psi(x,y,z=\pm L/2)=0.
\end{align}
 Since the translational symmetry is broken along the $z$-direction, we perform a Peierls 
substitution $k_z\to -\i\partial_z$ in the Schr{\"o}dinger equation $\hat{\mathcal{H}}\Psi = \varepsilon\Psi$. Assuming a 
plane-wave solution 
\begin{align}
\Psi \sim \e{\Lambda z},
\end{align}
 where $\Lambda$ determines whether the mode is evanescent or propagating, we 
solve the secular equation to find the allowed eigenvalues for $\Lambda$. These read $\Lambda=\pm\Lambda_\alpha$, where
\begin{align}\label{eq:lambda}
\Lambda_\alpha = [&A_1^2 - 2D_1(C-\varepsilon+D_2k_\perp^2) + 2B_1(B_2k_\perp^2-M) \notag\\
&+\alpha \sqrt{R}]^{1/2}/\sqrt{2(B_1^2-D_1^2)},\; \alpha=\pm1,
\end{align}
with the definition 
\begin{align}
R &= A_1^2[A_1^2 - 4D_1(C-\varepsilon+D_2k_\perp^2) - 4B_1(M-B_2k_\perp^2)]\notag\\
&-4A_2^2k_\perp^2(B_1^2-D_1^2)+4[B_1(C-\varepsilon+D_2k_\perp^2) \notag\\
&+ D_1(M-B_2k_\perp^2)]^2.
\end{align}
As demanded by consistency, Eq. (\ref{eq:lambda}) reduces to the 
 result of \cite{zhou_prl_08} in the limiting case of zero anisotropy and $C=0$. The total wavefunction $\Psi$ is a superposition 
 of the terms $\e{\pm\Lambda_\alpha z}$ with belonging normalization coefficients. The open boundary conditions at $z=\pm L/2$ 
 allow us to write down an implicit equation for the energy eigenvalues of $\Psi$. The values of $\varepsilon$ solving this 
 equation then correspond to the bulk states and, where possible, surface-bound states. We arrive at the following energy 
 eigenvalue equation:
\begin{align}\label{eq:boundstate}
\sum_\alpha \frac{t_\alpha}{t_{-\alpha}} = \frac{\Lambda_+^2+\Lambda_-^2 - (\Lambda_+^2-\Lambda_-^2)^2(B_1^2-D_1^2)/A_1^2}{\Lambda_+\Lambda_-},
\end{align}
where $t_\alpha = \tanh(\lambda_\alpha L/2)$. For arbitrary values of $k_\perp$, Eq. (\ref{eq:boundstate}) cannot be solved 
analytically. Instead, we employ a numerical solution of Eq. (\ref{eq:boundstate}) to find the allowed values of $\varepsilon$ 
for a given value of $k_\perp$. All other material parameters are specified in Eq. (\ref{eq:parameters}).

To highlight the dramatic difference between the surface states in Bi$_2$Se$_3$ and the edge states in HgTe, we show in 
Fig. \ref{fig:energyvsk} 
the energy dispersion as a function of the transverse momentum $k_\perp$. The plots for HgTe were obtained by utilizing 
the results in Ref.
 \cite{zhou_prl_08}. In the lower panel, it is seen how a gap $\Delta$ opens between the edge states at the $\Gamma$ point $k_\perp=0$ as 
 $L$ decreases. Let us emphasize here that the gap $\Delta$ between the edge/surface states is to be distinguished from the gap $|M|$ 
 between the bulk energy bands. For large $L \simeq 1000$ nm, the edge states remain ungapped for HgTe. In sharp contrast, the edge 
 states in Bi$_2$Se$_3$ \textit{remain completely ungapped in the entire regime} $L\in[150,1000]$ nm considered in Fig. \ref{fig:energyvsk}. 
 In fact, we find that a measurable gap $\Delta$ does not begin to open until widths of $L\leq 10$ nm are reached. This fact suggests that 
 much smaller samples that retain their conducting surface states can be fabricated in the case of Bi$_2$Se$_3$ than in the case of HgTe, 
 which is our first main result. Such an observation is crucial for the prospect of utilizing topological insulators such as Bi$_2$Se$_3$ 
 and HgTe in applications linked to quantum computing or spintronics. 

In order to understand the large quantitative difference between the necessary width $L$ that induces gapping between the surface/edge 
states in Bi$_2$Se$_3$ and HgTe, we study more carefully the eigenvalues $\Lambda_\alpha$. The physical interpretation of these 
quantities is that Re$\{\Lambda_\alpha\}$ corresponds to an inverse localization length (or, alternatively, penetration depth into 
the bulk) for the surface states. Therefore, the largest of the length scales (Re$\{\Lambda_+\})^{-1}$ and (Re$\{\Lambda_-\})^{-1}$ 
mainly determines the density profile for the surface states and their penetratation into the bulk. We have verified numerically 
that 
\begin{align}
(\text{Re}\{\Lambda_-\})^{-1} \geq (\text{Re}\{\Lambda_+\})^{-1}
\end{align}
for all energies inside the bulk gap $\pm |M|$ for both Bi$_2$Se$_3$ 
and HgTe, so that $\alpha=-1$ will determine the penetration depth of the surface states into the bulk. The penetration depth of 
the surface states can be estimated by 
\begin{align}
\xi = \hbar v_F/|M|,
\end{align} 
where the Fermi velocity is provided by $v_F=A/\hbar$ for HgTe and 
$v_F=A_2/\hbar$ for Bi$_2$Se$_3$. \cite{konig_science_07,zhang_nphys_09}. Since $A \simeq A_2$ while 
$|M|_\text{Bi$_2$Se$_3$} \gg |M|_\text{HgTe}$, the large difference in the distribution length of the surface/edge states 
stems from the sizable charge excitation gap $|M|$ in Bi$_2$Se$_3$. It is tempting to draw an analogy to the midgap Andreev-bound 
states in $d$-wave superconductors induced at the interface which extend a distance into the bulk proportional to the inverse 
of the superconducting gap \cite{hu_prl_94}. 

\begin{figure}[t!]
\centering
\resizebox{0.5\textwidth}{!}{
\includegraphics{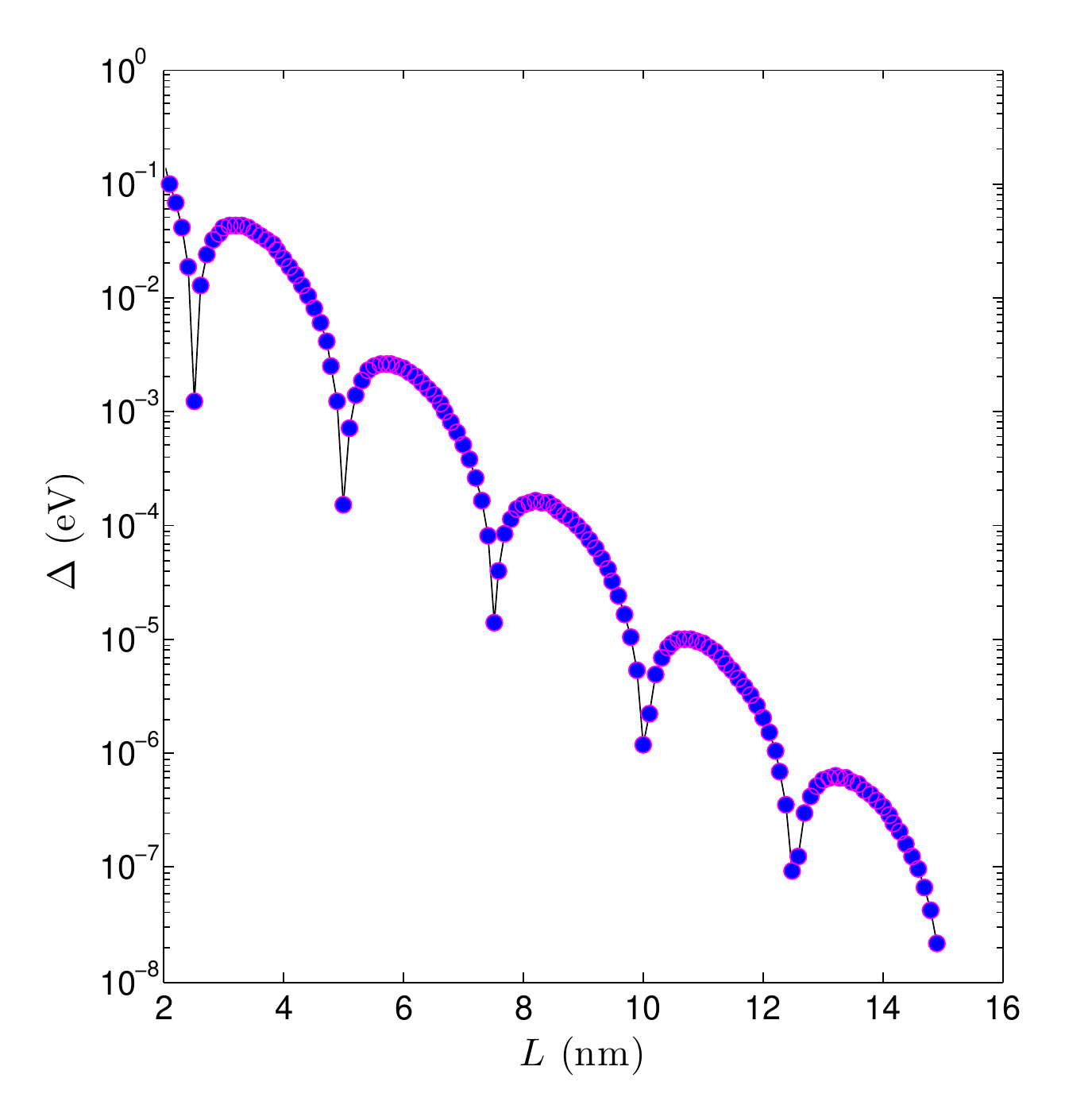}}
\caption{(Color online) Plot of the finite-size induced gap $\Delta$ between the surface states at $k_\perp=0$ for Bi$_2$Se$_3$ 
as a function of the width $L$. In contrast to HgTe where the decay with $L$ is purely exponential, an oscillatory pattern is 
superimposed on the decay for Bi$_2$Se$_3$. }
\label{fig:gap} 
\end{figure}

\begin{figure*}
\centering
\resizebox{0.85\textwidth}{!}{
\includegraphics{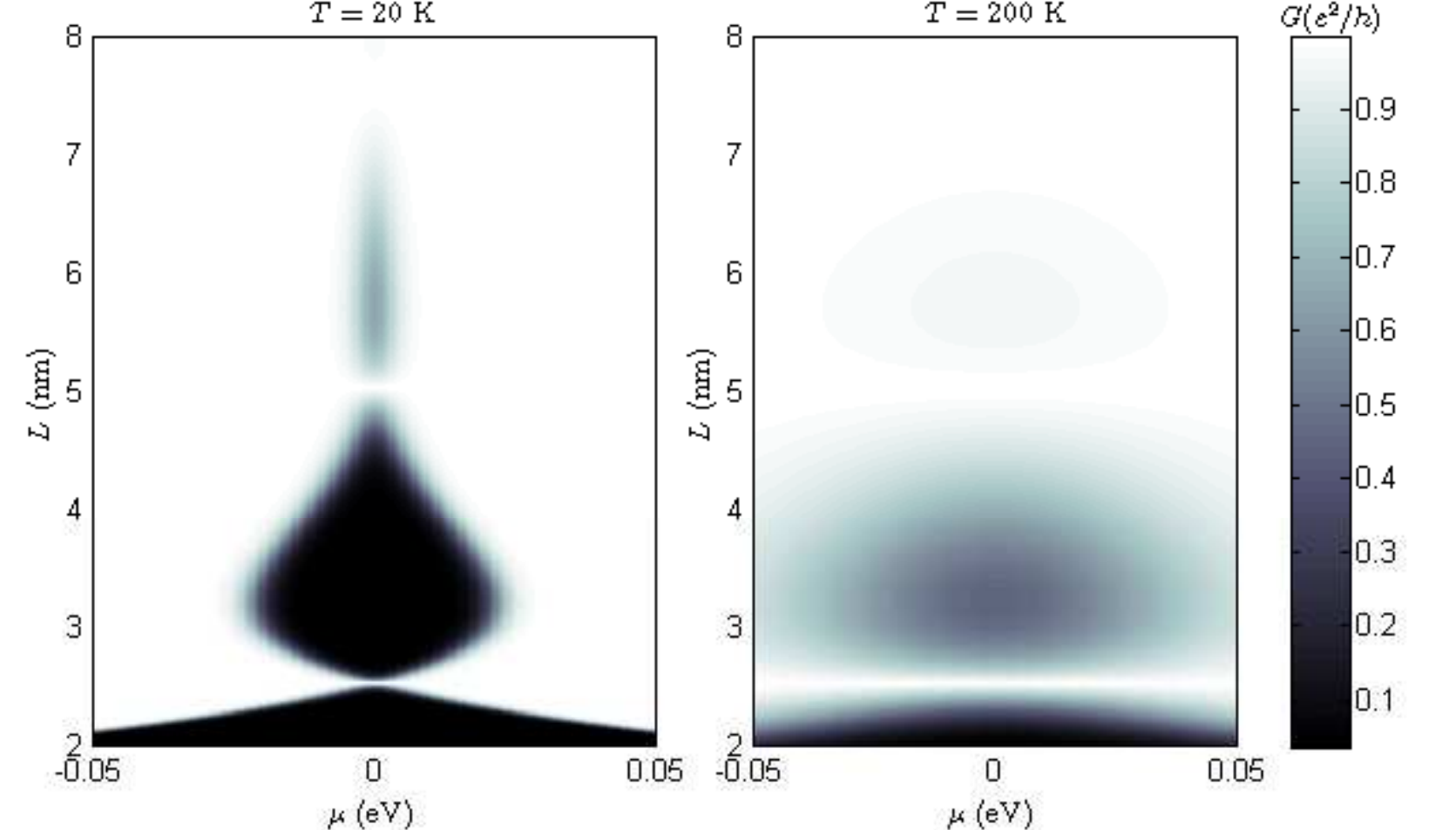}}
\caption{(Color online) Plot of the two-terminal conductance at $T=20$ K and $T=200$ K as a function of $\mu$ and $L$. }
\label{fig:cond} 
\end{figure*}

Our second main result is related to the manner in which the gap between the surface states in Bi$_2$Se$_3$ depends on the width $L$, 
in effect $\Delta=\Delta(L)$. It is instructive to first recall that $\Delta(L)$ in HgTe was shown to exhibit a purely exponential 
decay in Ref. \cite{zhou_prl_08}. To investigate the situation in Bi$_2$Se$_3$, we have solved numerically for $\Delta(L)$ and 
plotted the result in Fig. \ref{fig:gap}. It is seen that a qualitatively different scenario from HgTe transpires: the decay with 
$L$ is highly non-monotonous, and in fact features a superimposed oscillatory pattern on the exponential decay. The
experimental signature of such an oscillatory decay would be to
measure the conductance at a fixed chemical potential for several
samples with different widths $L$ and see how the conductance appears and
reappears, as we shall describe below.

We now proceed to explain the origin of the oscillatory decay of the gap $\Delta(L)$ found in Bi$_2$Se$_3$, considering the $\Gamma$ 
point $k_\perp=0$. The crucial observation in this context is that the eigenvalues $\Lambda_\alpha$ are \textit{not purely real} 
in the bulk insulating regime $\varepsilon \in [C-|M|, C+|M|]$. This is in contrast to HgTe, where $\Lambda_\alpha$ are purely 
real in this regime. For Bi$_2$Se$_3$, we find that 
\begin{align}
\Lambda_+ = \Lambda_{-}^* \text{ for } \varepsilon \in [C-|M|, C+|M|],
\end{align}
 as can be 
verified directly from Eqs. (\ref{eq:parameters}) and (\ref{eq:lambda}) since $R<0$. As a consequence, whereas a gap dependence 
of the type $\e{-\Lambda L}$ found for HgTe \cite{zhou_prl_08} dictates an exponential decay, it will give a superimposed 
oscillatory pattern on top of the exponential decay in the case of Bi$_2$Se$_3$. The natural question is then: is it possible 
to identify the reason for why $R<0$ in Bi$_2$Se$_3$ whereas $R>0$ in HgTe, leading to qualitatively different behavior of the 
surface states gap? Analyzing the expression for $R$ in Eq. (\ref{eq:lambda}) at $k_\perp=0$, it seen that neither the 3D 
nature or the anisotropy of the former material can be the reason, since there is no mixing between indices with subscript 
'1' and '2'. Therefore, $R<0$ seems to occur as a direct result of the material parameters given in Eq. (\ref{eq:parameters}). 
In principle, since all the parameters in the effective Hamiltonian of HgTe depend on the thickness $d$ of the sample (which also 
determines whether the material is in the trivial or topologically insulating state), it might be possible to obtain a conversion 
from exponential decay to oscillating decay also in HgTe, although this clearly warrants a separate investigation based on 
first-principles calculations.

The simplest way to experimentally detect both the gapping of the surface states \textit{and} their 
unusual dependence on $L$ is arguably a two-terminal geometry, as shown in Fig. \ref{fig:model}. In that case, the conductance can be 
evaluated in the Landauer-B{\"u}ttiker framework at a finite temperature $T$. Considering a zero-bias situation $eV\to0$, 
the transmission coefficient $\mathcal{T}$ may be written as 
\begin{align}
\mathcal{T}(\varepsilon) = N_c[\Theta(\varepsilon-\Delta/2) + \Theta(-\varepsilon-\Delta/2)],
\end{align} 
where $\Theta$ is the Heaviside-step function, $N_c$ is the number of conducting 
channels on the surfaces $z=\pm L/2$, and $\varepsilon$ is the quasiparticle energy. Similarly to Ref. \cite{zhou_prl_08}, one arrives at the following expression for the 
conductance normalized against $N_c$: 
\begin{align}
G = \frac{e^2}{h}\{1+[1+\e{\beta(\Delta/2 -\mu)}]^{-1} - [1+\e{\beta(-\Delta/2 -\mu)}]^{-1}\},
\end{align} 
with $\beta^{-1} = k_BT$. As a direct consequence of the gap $\Delta$, the conductance is suppressed as $T\to0$. In Fig. \ref{fig:cond}, 
we plot the conductance as a function of $L$ and the chemical potential $\mu$, comparing the temperatures $T=20$ K and $T=200$ K. As seen, the conductance 
displays oscillations for a fixed $\mu$, and the oscillation length depends on the value of the chemical potential. This is 
qualitatively completely different from HgTe, where the conductance displays a monotonic dependence on $L$. The oscillatory 
features become more smeared at elevated temperatures, as expected.

Finally, we note that after this paper was submitted for publication, we 
learned about the very recent work of H.-Z. Lu \etal \cite{lu_arxiv_09} and C.-X. Liu \etal \cite{liu_arxiv_09}, in which similar finite size effects have been predicted.

\section{Summary}

In summary, we have investigated finite size effects on the surface states in the strong topological insulator Bi$_2$Se$_3$, 
comparing the results also with those recently reported for HgTe \cite{zhou_prl_08}. We demonstrate that the surface states 
respond differently to finite size effects in these materials, both quantitatively and qualitatively. First of all, while 
the edge states become measurably gapped around $L\simeq 200$ nm in HgTe \cite{zhou_prl_08}, the surface-states in Bi$_2$Se$_3$ 
display a considerable robustness towards decreasing $L$ and become measurably gapped around $L \simeq10$ nm. In this way, 
the topological surface state remains intact for a wider range of widths $L$. Secondly, the gapping between the surface states features 
a qualitatively distinct dependence on $L$ in Bi$_2$Se$_3$ compared to HgTe, namely an oscillatory decay with $L$ which stems 
from the material parameters that give an eigenvalue pair $\Lambda_\alpha$ that are complex conjugates. Both of these effects 
can be experimentally detected in a two-terminal geometry by varying the width and  the chemical potential of the junction.

\acknowledgments 

J.L. and A.S. were supported by the Research Council of Norway, 
Grants No. 158518/431 and No. 158547/431 (NANOMAT), and Grant No. 167498/V30 (STORFORSK). T.Y. acknowledges support by JSPS.

\end{document}